\documentclass{article}
\usepackage{amssymb}

\usepackage{amsmath}

\input{tcilatex}

\begin{document}

\title{Three Different Sizes Obtained Using Light Scattering Techniques }
\author{Yong Sun}
\maketitle

\begin{abstract}
The average scattered intensity is determined by the optical characteristics
of particles in dispersion. The normalized time auto-correlation function of
the scattered light intensity $g^{\left( 2\right) }\left( \tau \right) $
includes both the optical and hydrodynamic information of particles. From
the different characteristics of particles, the particle sizes can be
obtained. In this paper, three sizes: the static radius $R_{s}$, the
hydrodynamic radius $R_{h}$ and the apparent hydrodynamic radius $R_{h,app}$%
\ are discussed using dilute water dispersions of homogenous spherical
particles poly($N$-isopropylacrylamide) microgels, with a simple assumption
that the hydrodynamic radius is in proportion to the static radius, when
Rayleigh-Gans-Debye approximation is valid. Our results show that the
expected values of the normalized time auto-correlation function of the
scattered light intensity are consistent with the experimental data very
well and the difference between the static radius and the apparent
hydrodynamic radius is large and the difference between the hydrodynamic
radius and the apparent hydrodynamic radius is influenced by the particle
size distribution.
\end{abstract}

\section{Introduction}

A great deal of information about particles in dispersion can be obtained
using light scattering techniques. One of the main applications of the light
scattering techniques is that measure the particle sizes. The static light
scattering technique $\left( SLS\right) $ obtains the size information from
the optical characteristics and the dynamic light scattering technique $%
\left( DLS\right) $ obtains the size information from both the optical and
hydrodynamic features of particles.

During the last few decades, the standard method of cumulant is used to
obtain the apparent hydrodynamic radius $R_{n,app}$ and its distribution $%
G\left( R_{h,app}\right) $ for particles from the normalized time
auto-correlation function of the scattered light intensity $g^{\left(
2\right) }\left( \tau \right) $ with the assistance of the Einstein-Stokes
relation, where $\tau $ is a delay time. The treatment of SLS is simplified
to the Zimm plot, Berry plot or Guinier plot etc. to obtain the root
mean-square radius of gyration $\left\langle R_{g}^{2}\right\rangle ^{1/2}$
and the molar mass of particles provided that the particle sizes are small.
In order to obtain more accurate information about the particles, people
have explored the relationships between the physical quantities obtained
using SLS and DLS techniques. People $\left[ \cite{re1}-\cite{re3}\right] $
think that the measurements of the dimensionless shape parameter $\rho
=\left\langle R_{g}^{2}\right\rangle ^{1/2}/R_{h,app}$ can provide a
relatively unambiguous test for the particle shape. In this judgement, there
have an assumption that the sizes obtained from SLS and DLS are the same.
However, the sizes obtained from SLS and DLS are different physical
quantities $\left[ \cite{re4},\cite{re5}\right] $. Exactly using the light
scattering techniques, three different sizes can be obtained for homogenous
spherical particles: one is static radii $R_{s}$ obtained from the optical
characteristics; the second is hydrodynamic radii $R_{h}$ obtained from the
hydrodynamic features and the third is apparent hydrodynamic radii $%
R_{h,app} $ determined by the both optical and hydrodynamic characteristics
of particles.

In this article, the three different sizes will be discussed using PNIPAM
microgel samples with a simple assumption that hydrodynamic radius $R_{h}$
is in proportion to the static radius $R_{s}$. With this simple assumption,
the relationship between the SLS\ and DLS can be built for homogenous
spherical particles. Our results show that the expected values of the
normalized time auto-correlation function of the scattered light intensity
are consistent with the experimental data very well and the difference
between the static radius and the apparent hydrodynamic radius is large and
the difference between the hydrodynamic radius and the apparent hydrodynamic
radius is influenced by the particle size distribution.

\section{Theory}

For homogeneous spherical particles where the Rayleigh-Gans-Debye $\left(
RGD\right) $ approximation is valid, the normalized time auto-correlation
function of the electric field of the scattered light $g^{\left( 1\right)
}\left( \tau \right) $ is given by

\begin{equation}
g^{\left( 1\right) }\left( \tau \right) =\frac{\int_{0}^{\infty
}R_{s}^{6}P\left( q,R_{s}\right) G\left( R_{s}\right) \exp \left(
-q^{2}D\tau \right) dR_{s}}{\int_{0}^{\infty }R_{s}^{6}P\left(
q,R_{s}\right) G\left( R_{s}\right) dR_{s}},  \label{Grhrs}
\end{equation}
where $q$ is the scattering vector, $R_{s}$ is the static radius, $\tau $ is
the delay time, $D$ is the diffusion coefficient, $G\left( R_{s}\right) $ is
the number distribution and the form factor $P\left( q,R_{s}\right) $ is

\begin{equation}
P\left( q,R_{s}\right) =\frac{9}{q^{6}R_{s}^{6}}\left( \sin \left(
qR_{s}\right) -qR_{s}\cos \left( qR_{s}\right) \right) ^{2}.  \label{factor}
\end{equation}
In this discussion, the number distribution is chosen as a Gaussian
distribution

\begin{equation*}
G\left( R_{s};\left\langle R_{s}\right\rangle ,\sigma \right) =\frac{1}{%
\sigma \sqrt{2\pi }}\exp \left( -\frac{1}{2}\left( \frac{R_{s}-\left\langle
R_{s}\right\rangle }{\sigma }\right) ^{2}\right) ,
\end{equation*}
where $\left\langle R_{s}\right\rangle $ is the mean static radius and $%
\sigma $ is the standard deviation relative to the mean static radius.

From the Stokes-Einstein relation

\begin{equation}
D=\frac{k_{B}T}{6\pi \eta _{0}R_{h}},
\end{equation}
where $\eta _{0}$, $k_{B}$, $T$ and $R_{h}$ are the viscosity of the
solvent, Boltzmann's constant, absolute temperature and hydrodynamic radius
of a particle, the hydrodynamic radius can be obtained.

For simplicity, we assume that the relationship between the static and
hydrodynamic radii can be written as 
\begin{equation}
R_{h}=aR_{s},  \label{RsRh}
\end{equation}
where $a$ is a constant. With the function between the normalized time
auto-correlation function of the scattered light intensity $g^{\left(
2\right) }\left( \tau \right) $ and the normalized time auto-correlation
function of the electric field of the scattered light $g^{\left( 1\right)
}\left( \tau \right) $ \cite{re6}

\begin{equation}
g^{\left( 2\right) }\left( \tau \right) =1+\beta \left( g^{\left( 1\right)
}\right) ^{2},  \label{G1G2}
\end{equation}
the relationship between the static and dynamic light scattering is built
and the values of the normalized time auto-correlation function of the
scattered light intensity $g^{\left( 2\right) }\left( \tau \right) $ can be
expected using the size information obtained from SLS.

If the first cumulant is used to obtain the apparent hydrodynamic radius $%
R_{h,app}$, the constant $a$ can be determined approximately using

\begin{equation}
a=\frac{R_{h,app}\int_{0}^{\infty }R_{s}^{6}P\left( q,R_{s}\right) G\left(
R_{s}\right) dR_{s}}{\int_{0}^{\infty }R_{s}^{7}P\left( q,R_{s}\right)
G\left( R_{s}\right) dR_{s}}.  \label{cona}
\end{equation}

\section{Experiment}

The SLS and DLS spectroscopies were measured using the instrument built by
ALV-Laser Vertriebsgesellschaft m.b.H (Langen, Germany). It utilizes an
ALV-5000 Multiple Tau Digital Correlator and a JDS Uniphase 1145P\ He-Ne
laser to provide a 23 mW vertically polarized laser\ at wavelength of 632.8
nm.

In this experiment, $N$-isopropylacrylamide (NIPAM, monomer) from Acros
Organics was recrystallized from hexane/acetone solution. Potassium
persulfate (KPS, initiator) and $N,N^{\prime }$-methylenebisacrylamide (BIS,
cross-linker) from Aldrich were used as received. Fresh de-ionized water
from a Milli-Q Plus water purification system (Millipore, Bedford, with a
0.2 $\mu m$ filter) was used throughout the whole experiment. The synthesis
of gel particles was described elsewhere $\left[ \cite{re7},\cite{re8}\right]
$ and the recipes of the\ batches used in this work are listed in Table 1.

\begin{center}
$\underset{\text{Table 1. Synthesis conditions for PNIPAM particles.}}{
\begin{tabular}{|c|c|c|c|c|c|}
\hline
Sample & $T\left( ^{o}C\right) $ & $t\left( hrs\right) $ & $%
W_{N}+W_{B}\left( g\right) $ & $KPS\left( mg\right) $ & $n_{B}/n_{N}$ \\ 
\hline
$PNIPAM-0$ & $70\pm 1$ & $4.0$ & $1.00$ & $40$ & $0$ \\ \hline
$PNIPAM-1$ & $70\pm 1$ & $4.0$ & $1.00$ & $40$ & $1.0\%$ \\ \hline
$PNIPAM-2$ & $70\pm 1$ & $4.0$ & $1.00$ & $40$ & $2.0\%$ \\ \hline
$PNIPAM-5$ & $70\pm 1$ & $4.0$ & $1.00$ & $40$ & $5.0\%$ \\ \hline
\end{tabular}
}$
\end{center}

The four samples were named according to the molar ratios $n_{B}/n_{N}$ of $%
N,N^{\prime }$-methylenebisacrylamide over $N$-isopropylacrylamide. They
were centrifuged at 14,500 RPM followed by decantation of the supernatants
and re-dispersion in fresh de-ionized water four times to remove of free
ions and any possible linear chains. Then the samples were diluted for light
scattering to weight factors of $5.9\times 10^{-6}$, $8.56\times 10^{-6}$, $%
9.99\times 10^{-6}$ and $8.38\times 10^{-6}$ for $PNIPAM-0$, $PNIPAM-1$, $%
PNIPAM-2$ and $PNIPAM-5$\ respectively. Before\ the measurements were made,
0.45 $\mu m$ filters (Millipore, Bedford) were used to do dust free for the
samples $PNIPAM-1,$ $PNIPAM-2$ and $PNIPAM-5$. Measurements of the
normalized time auto-correlation function of the scattered light intensity $%
g^{\left( 2\right) }\left( \tau \right) $ were performed five times at each
angle for all samples except the $PNIPAM-0$ sample, in which measurements
were performed only twice.

\section{Data Analysis}

The mean static radius $\left\langle R_{s}\right\rangle $ and the standard
deviation $\sigma $ of the PNIPAM microgel samples are obtained from the SLS
data \cite{re9}. For $PNIPAM-1,$ the mean static radius $\left\langle
R_{s}\right\rangle $, the standard deviation $\sigma $ and $\chi ^{2}$\
obtained at different temperatures are listed in Table 4.1.

\begin{center}
$\underset{\text{Table 4.1 The fit results obtained from SLS for }PNIPAM-1%
\text{ at different temperatures.}}{
\begin{tabular}{|c|c|c|c|}
\hline
Temperature ($^{o}C$) & $\left\langle R_{s}\right\rangle (nm)$ & $\sigma
(nm) $ & $\chi ^{2}$ \\ \hline
25 & 277.7$\pm $0.5 & 23.1$\pm $0.9 & 1.84 \\ \hline
27 & 267.1$\pm $0.5 & 23.1$\pm $0.8 & 2.50 \\ \hline
29 & 254.3$\pm $0.1 & 21.5$\pm $0.3 & 2.15 \\ \hline
31 & 224.8$\pm $0.9 & 30.6$\pm $0.9 & 3.31 \\ \hline
33 & 119.9$\pm 0.9$ & 19.8$\pm $0.6 & 3.16 \\ \hline
36 & 110.4$\pm $0.9 & 17.3$\pm $0.7 & 4.19 \\ \hline
40 & 111.7$\pm $0.9 & 14.8$\pm $0.8 & 2.73 \\ \hline
\end{tabular}
\ \ }$
\end{center}

The values of the apparent hydrodynamic radius at different scattering
angles were obtained using the first cumulant analysis $\left[ \cite{re10}- 
\cite{re12}\right] $. In order to avoid the consideration for the large
values of $\chi ^{2}$, all the fit results obtained using the first cumulant
analysis are chosen under this condition $\chi ^{2}\leqslant 2$. For $%
PNIPAM-1$ at a temperature of 27$^{o}C$, the values of the apparent dynamic
radius at different scattering angles are list in Table 4.2. The ratios of
the hydrodynamic radius over the static radius calculated using Eq. \ref
{cona} at different scattering angles also are listed in Table 4.2.

\begin{center}
$\underset{
\begin{array}{c}
\text{Table 4.2 The values of apparent hydrodynamic radius and constant }a
\\ 
\text{\ for }PNIPAM-1\text{ at different scattering angles and a temperature
of }27^{o}C\text{.}
\end{array}
}{
\begin{tabular}{|c|c|c|}
\hline
Scattering Angle $\left( Degree\right) $ & $R_{h,app}\left( nm\right) $ & $a$
\\ \hline
30 & 329.$\pm 4$. & 1.19 \\ \hline
35 & 331.1$\pm $0.7 & 1.21 \\ \hline
40 & 329.6$\pm $0.9 & 1.21 \\ \hline
45 & 329.8$\pm $0.5 & 1.22 \\ \hline
50 & 329.$\pm 1$. & 1.22 \\ \hline
55 & 326.4$\pm $0.2 & 1.23 \\ \hline
60 & 327.$\pm $2. & 1.25 \\ \hline
65 & 323.$\pm $2. & 1.26 \\ \hline
\end{tabular}
\ }$
\end{center}

From the results shown in Table 4.2, the value of $a$ almost is a constant.
If the value of constant $a$ was set to 1.21, the expected values of $%
g^{\left( 2\right) }\left( \tau \right) $ calculated using the mean static
radius $\left\langle R_{s}\right\rangle $, the standard deviation $\sigma $
and Eqs. \ref{Grhrs} and \ref{G1G2} were compared with the experimental data
measured at scattering angles 30$^{o}$, 45$^{o}$ and 60$^{o}$, respectively.
The results are shown in Fig. 4.1. The results show that the expected values
are consistent with the experimental data very well. With the constant $a$:
1.21, the mean hydrodynamic radius $323.2\pm 0.6$ $nm$ can be obtained and
this value only has a little difference with the values of the apparent
hydrodynamic radii obtained using the first cumulant at different scattering
angles.

From the results shown in Table 4.1, at a temperature of 33$^{o}C,$ the
particle size distribution of the $PNIPAM-1$ sample is wide. The values of
the apparent hydrodynamic radius are listed in Table 4.3. The ratios of the
hydrodynamic radius over the static radius calculated using Eq. \ref{cona}
also are listed in Table 4.3. The value of $a$ still almost is a constant.
If the value of constant $a$ was set to 1.55, the expected values of $%
g^{\left( 2\right) }\left( \tau \right) $ calculated using the mean static
radius $\left\langle R_{s}\right\rangle $, the standard deviation $\sigma $
and Eqs. \ref{Grhrs} and \ref{G1G2} were compared with the experimental data
measured at scattering angles 30$^{o}$, 60$^{o}$ and 90$^{o}$, respectively.
The results are shown in Fig. 4.2. The results show that the expected values
are consistent with the experimental data very well. With the constant $a$:
1.55, the mean hydrodynamic radius $186.\pm 1$. $nm$ was obtained and the
difference between the mean hydrodynamic radius and the apparent
hydrodynamic radius is large.

\begin{center}
$\underset{
\begin{array}{c}
\text{Table 4.3 The values of apparent hydrodynamic radius and constant }a
\\ 
\text{ for }PNIPAM-1\text{ at different scattering angles and a temperature
of }33^{o}C\text{.}
\end{array}
}{
\begin{tabular}{|c|c|c|}
\hline
Scattering Angle $\left( Degree\right) $ & $R_{h,app}\left( nm\right) $ & $a$
\\ \hline
30 & 212.1$\pm $0.6 & 1.55 \\ \hline
35 & 208.3$\pm $0.2 & 1.53 \\ \hline
40 & 208.7$\pm $0.6 & 1.54 \\ \hline
45 & 207.3$\pm $0.3 & 1.53 \\ \hline
50 & 206.7$\pm $0.4 & 1.53 \\ \hline
55 & 206.$\pm $1. & 1.53 \\ \hline
60 & 205.$\pm $1. & 1.53 \\ \hline
65 & 204.9$\pm $0.6 & 1.54 \\ \hline
70 & 205.$\pm $1. & 1.55 \\ \hline
75 & 205.$\pm 1.$ & 1.56 \\ \hline
80 & 205.$\pm $1. & 1.56 \\ \hline
85 & 205.$\pm $1. & 1.57 \\ \hline
90 & 203.2$\pm $0.7 & 1.57 \\ \hline
95 & 204.$\pm $1. & 1.58 \\ \hline
\end{tabular}
\ }$
\end{center}

\section{Results and Discussion}

From the above analysis, three different sizes can be obtained using the
light scattering techniques. In general, the values of constant $a$ vary
little during the small scattering angle range and the values of the
apparent hydrodynamic radius $R_{h,app}$ are a function of the scattering
angle. In order to compare the three different sizes conveniently, all the
mean hydrodynamic and the apparent hydrodynamic radii were obtained at a
scattering angle of 30$^{o}$. All three sizes obtained at different
temperatures for $PNIPAM-1$ are shown in Fig. 5.1. The picture shows that
the difference between the mean static radii and the apparent hydrodynamic
radii is large and the difference between the mean hydrodynamic radii and
the apparent hydrodynamic radii is influenced by the particle size
distribution.

The three different sizes represent the different characteristics of
particles. When temperature change from 25$^{o}C$ to 40$^{o}C$, the
characteristics of PNIPAM\ microgel samples change from being hydrophilic to
hydrophobic. It is possible that this change makes the different effects on
the optical and hydrodynamic characteristics of particles. In order to show
the results of this effects, the ratios $R_{h,app}^{T}/\left\langle
R_{s}^{T}\right\rangle $ and $\left\langle R_{h}^{T}\right\rangle
/\left\langle R_{s}^{T}\right\rangle $ as a function of temperature $T$ are
shown in Figs. 5.2.a and 5.2.b, respectively.

From the fit results for the four PNIPAM microgel sample, the particle size
distributions are narrow both below and above the phase transition and are
wide near the phase transition. Figure 5.2 shows clearly that the difference
between the apparent hydrodynamic radii and the mean hydrodynamic radii is
influenced by the distribution widths. For small poly-dispersities, the
value of apparent hydrodynamic radius is almost the same as that of the mean
hydrodynamic radius. For wide distributions, the part of apparent
hydrodynamic radius represents the effects of the particle size
distribution. From the theoretical analysis of cumulant, the apparent
hydrodynamic radius is obtained from the average of the term $\exp \left(
-q^{2}D\tau \right) $ in distribution $G\left( R_{s}\right) $ with the
weight $R_{s}^{6}P\left( q,R_{s}\right) $, where $\exp \left( -q^{2}D\tau
\right) $ represents the hydrodynamic features of particles. For the
mono-disperse particle systems, since the effects of scattered intensity are
cancelled, the apparent hydrodynamic radius is equal to the hydrodynamic
radius. For polydisperse particle systems, the apparent hydrodynamic radius
shows the total effects of the optical and hydrodynamic characteristics of
particles.

Since the PNIPAM microgels possess the temperature sensitivity during the
temperature range $15^{o}C-50^{o}C$, a few authors $\left[ \cite{re13},\cite
{re14}\right] $used the equilibrium swelling ratios $%
R_{h,app}^{T}/R_{h,app}^{T_{0}}$ to show the volume phase transition. For
our samples, the volume phase transition during the temperature range $%
25^{o}C-40^{o}C$ is shown in Fig. 5.3 using the equilibrium swelling ratios
of the mean static radii and the apparent hydrodynamic radii, respectively.
All radii are compared to that measured at a temperature of 40$^{o}C$. The
ratios of $\left\langle R_{s}^{T}\right\rangle /\left\langle
R_{s}^{40^{o}C}\right\rangle $ and $R_{h,app}^{T}/R_{h,app}^{40^{o}C}$ are
shown in Fig. 5.3.a and b respectively.

From the chemical knowledge, the materials of PNIPAM possess the temperature
sensitivity. If adding the $N,N^{\prime }$-methylenebisacrylamide, the
temperature sensitivity of PNIPAM microgels will be influenced by the
content of the $N,N^{\prime }$-methylenebisacrylamide which does not possess
the temperature sensitivity. If the content of the $N,N^{\prime }$%
-methylenebisacrylamide continues to increase, the temperature sensitivity
of PNIPAM microgels is becoming weak. Figure 5.3 clearly shows the feature.
The phase transition of PNIPAM microgels, indicated as the ratios $%
\left\langle R_{s}^{T}\right\rangle /\left\langle
R_{s}^{40^{o}C}\right\rangle $ or $R_{h,app}^{T}/R_{h,app}^{40^{o}C}$ as a
function of $T$, becomes less sharp and occurs in a broader $T$ range as the 
$N,N^{\prime }$-methylenebisacrylamide content is increased.

\section{Conclusion}

The average scattered intensity is determined by the optical characteristics
of particles in dispersion. The normalized time auto-correlation function of
the scattered light intensity $g^{\left( 2\right) }\left( \tau \right) $
includes both the optical and hydrodynamic information of particles. Using
the light scattering techniques, three different particle sizes can be
measured. The static radius represents the optical characteristics, the
hydrodynamic radius shows the hydrodynamic features and the apparent
hydrodynamic radius represents both the optical and hydrodynamic
characteristics of particles. For narrow distributions, the value of the
apparent hydrodynamic radius is a good approximation of the mean
hydrodynamic radius. In general, the sizes obtained from SLS and DLS are
different. If the relationship between the optical and hydrodynamic features
of particles can be understood, the accurate relationship between the SLS\
and DLS can be built. Then the static sizes also can be obtained from the
DLS data. The accurate relationship between the optical and hydrodynamic
features of particles can be further explored.

Fig. 4.1 The expected and experimental values of the normalized time
auto-correlation function of the scattered light intensity $g^{\left(
2\right) }\left( \tau \right) $ for $PNIPAM-1$ at a temperature of $27^{o}C$%
. The symbols show the experimental results and the line shows the
calculated values with the simple assumption $R_{h}=1.21R_{s}$.

Fig. 4.2 The expected and experimental values of the normalized time
auto-correlation function of the scattered light intensity $g^{\left(
2\right) }\left( \tau \right) $ for $PNIPAM-1$ at a temperature of $33^{o}C$%
. The symbols show the experimental results and the line shows the
calculated values with the simple assumption $R_{h}=1.55R_{s}$.

Fig. 5.1 Values of the apparent hydrodynamic radii $\left( \blacklozenge
\right) $ , the mean hydrodynamic radii $\left( \blacktriangle \right) $ at
a scattering angle of $30^{o}$ and the mean static radii $\left( \bullet
\right) $ for $PNIPAM-1$ at different temperatures.

Fig. 5.2 The comparisons for three different sizes obtained using light
scattering technique. a). The ratios $R_{h,app}^{T}/\left\langle
R_{s}^{T}\right\rangle $ of the apparent hydrodynamic over the mean static
radii are shown. b). The ratios $\left\langle R_{h}^{T}\right\rangle
/\left\langle R_{s}^{T}\right\rangle $ of the mean hydrodynamic over the
mean static radii are shown for $PNIPAM-0$, $PNIPAM-1$, $PNIPAM-2$ and $%
PNIPAM-5$ during the temperature range from $25^{o}C$ to $40^{o}C$.

Fig. 5.3 \ The volume phase transition of $PNIPAM-0$, $PNIPAM-1$, $PNIPAM-2$
and $PNIPAM-5$. a). The phase transition was shown using the ratios of the
mean static radii $\left\langle R_{s}^{T}\right\rangle $ at temperature $T$
\ to that $\left\langle R_{s}^{40^{o}C}\right\rangle $ at $40^{o}C.$ b). The
phase transition was shown using the ratios of the apparent hydrodynamic
radii $R_{h,app}^{T}$ at temperature $T$ to that $R_{h,app}^{40^{o}C}$ at $%
40^{o}C$.

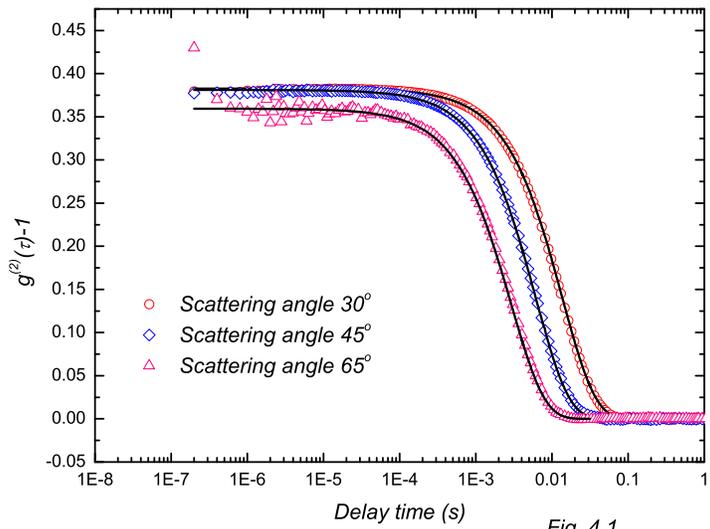
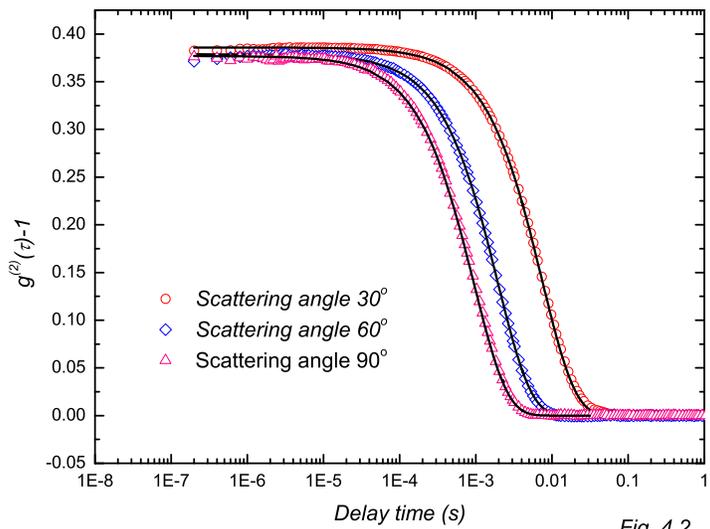
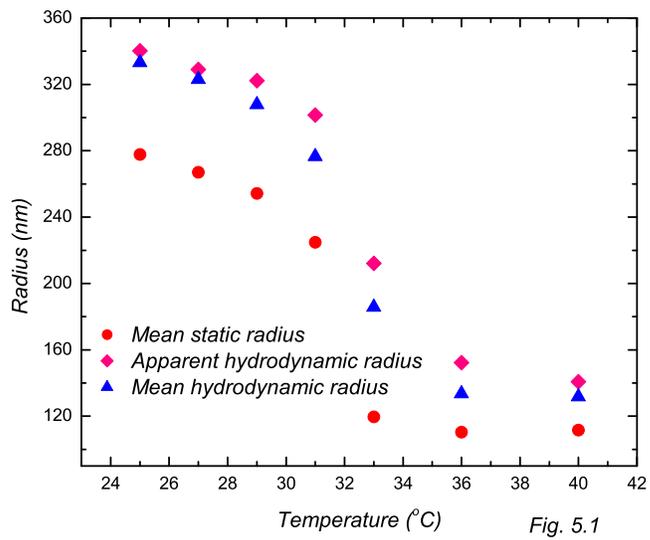
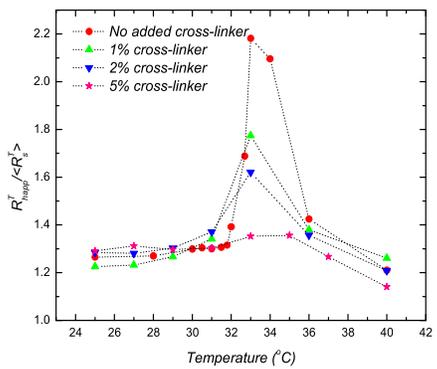
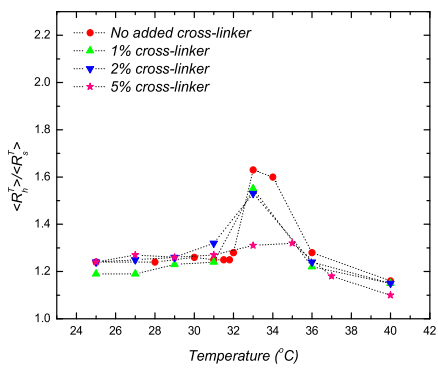
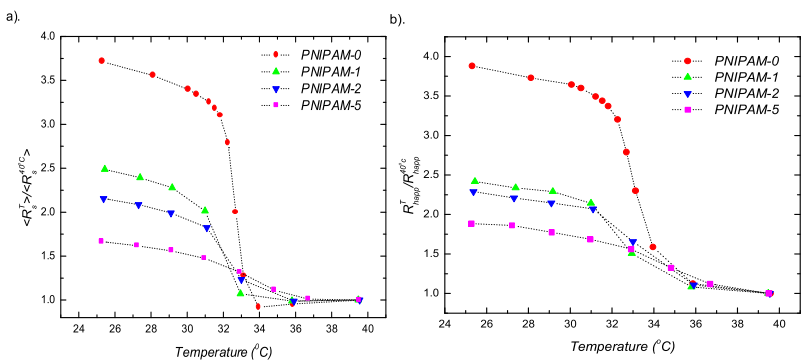
\end{document}